# Incorporating demand response of electric vehicles in scheduling of isolated microgrids with renewables using a bi-level programming approach

**Yang Li, Senior Member, IEEE, Kang Li**

School of Electrical Engineering, Northeast Electric Power University, Jilin 132012, China

Corresponding author: Yang Li (e-mail: liyang@neepu.edu.cn).

This work was supported in part by the China Scholarship Council (CSC) under Grant 201608220144.

**ABSTRACT** In this work, a novel optimal scheduling approach is proposed for isolated microgrids (MGs) with renewable generations by incorporating demand response of electric vehicles (EVs). First, a bi-level programming-based MG scheduling model is proposed under real-time pricing environments, where the upper- and lower- levels seek to minimize the MG net operating cost and the EV charging cost. Second, a hybrid solution algorithm called JAYA-interior point method is put forward to solve the model. And finally, the simulation results demonstrate that incorporating demand response of electric vehicles is able to guide EV users to actively participate in MG scheduling and achieve the peak load shaving, which offers a fundamental way to balance the interests between MG and EV users.

## I. INTRODUCTION
### A. MOTIVATION
With the increasingly serious energy crisis and environmental problems, it is has become a broad consensus to research and leverage renewable generations and electric vehicles [1, 2]. At the same time, as an effective carrier of distributed renewable generations, a microgrid (MG) can fully promote the integration of renewable generations and has become an important part of smart grids [3, 4]. Compared with the grid-connected microgrids, isolated microgrids (IMGs) have obvious advantages in areas which are inaccessible to the main grid, such as islands, remote areas, deserts, etc. [5, 6]. However, the inherent uncertainty of renewable generations is difficult to guarantee operational reliability and power supply security, especially for IMGs. Consequently, it is a tricky problem to address renewable uncertainty in IMGs.

Recent studies have shown that the integration of electric vehicles (EVs) into MGs via vehicle-to-grid (V2G) is beneficial to implement energy conservation and emission reduction [7], but the EV charging behaviors have strong time and space uncertainties [8]. Therefore, the integration of EVs will further increase the difficulty of MG scheduling. As a new type of demand-side management strategy, price-based demand response (PBDR) mainly includes real-time prices and time-of-use tariffs, which is able to maintain supply-demand balance by flexibly adjusting the load. In addition, demand response can effectively consume uncertain renewable generations including PV and WT. Therefore, how to comprehensively consider EV demand response and uncertain renewable generations is an urgent and challenging problem.

### B. LITERATURE REVIEW
Optimal scheduling of microgrids with renewable uncertainties has always been an important issue to ensure the MG safe and economic operation. Compared with a traditional power system, MGs generally include a large proportion of renewable generations such as wind turbines (WT) and photovoltaic (PV). At present, the accuracy of the prediction of renewable outputs needs to be further improved, and the fluctuations of renewable outputs have a great influence on the safe and economic operation of MGs [9-12]. At present, many scholars have performed a series of studies on the economic dispatch of MGs [13-16]. Reference [13] develops a linear programming cost minimization model for MGs. In [14], an economic dispatch model is proposed for microgrids, which considers simultaneously cogeneration and reserve capacity constraints. In order to address the uncertainties of renewable generations, reference [15] uses a risk metric called conditional value-at-risk, which aims to limit the possibility of the shortage of renewable generations at a certain confidence level. Reference [16] adopts the probabilistic model to simulate the renewable outputs, and proposes a multi-objective MG scheduling model that combines energy storage and user experience.

With the popularization of EVs, their load demands have an important impact on the operation of power grids. Recently, vehicle-to-grid (V2G) related studies are becoming a hot topic. (1) regarding EV charging cost: in [17], by considering the relationship between the acceptable charging power and the state of charge (SOC) of EV batteries, a heuristic method is proposed for minimizing the EV charging cost; to minimize the EV charging cost, a strategy is presented for coordinating autonomous plug-in EV charging using the concept of non-cooperative games [18]. (2) In terms of V2G technology: a real-time pricing mechanism is developed to consider the dynamic relationships between supply and demand in an MG with battery swapping stations in [19]; reference [20] proposes an optimal V2G aggregator for frequency regulation; reference [21] builds a market model to enable agents to participate in bilateral transactions and optimizes the EV charging process via dynamic allocation; in [22], the impact of charging hybrid EVs on a residential distribution grid has been analyzed; a new dynamic charging approach is proposed for electric vehicle fleets by using adaptive learning in [23].

Based on the above existing works, basic scheduling issues have been resolved. Unfortunately, there are still some gaps in MG scheduling. To the best knowledge of authors, there are only a few related investigations that utilize EVs to participate in demand response in MG scheduling. In [24], by coordinating demand response and storage, a new emergency operation approach is developed with consideration of the responsiveness of EVs and load. Reference [25] presents price-based and incentive-based demand response strategies and examines their impact on the MG economy under grid-connected and isolated modes. Obviously, it is of great significance for MG optimal scheduling to comprehensively considerate both renewable uncertainties and EV demand response, but until now, there are few investigations reported for addressing this problem. In view of this, this work proposes a novel bi-level programming-based scheduling approach for MGs with renewables by incorporating EV demand response. Note that this problem is nonlinear and nonconvex. Meanwhile, the associated problems, such as getting trapped in local minima, and scalability to high-number of variables, have always existed. Recent studies demonstrate that the hybrid analysis-heuristic solution approach is an effective way to address these problems [19, 26].

### C. CONTRIBUTIONS

The main contributions of this paper are as follows: (1) A bi-level programming-based MG scheduling model under real-time pricing environments is proposed, which is the first attempt to coordinate renewable generation uncertainty and EV demand response; (2) A new hybrid solution algorithm called JAYA-IPM is developed to solve the model with sufficient optimality and high computation efficiency; (3) And finally, the simulation results indicate that demand response of electric vehicles is able to guide EV users to actively participate in MG scheduling and achieve the peak load shaving, which provides a fundamental way to balance the interests of both MG and users.

The novelty of this study lies primarily in our attempt to propose a new method that can coordinate the EV demand response and renewable generation uncertainty in MG scheduling.

### D. PAPER ORGANIZATION

The rest of this paper is organized as follows. Section II gives the modeling of MG and EV. In Section III, the problem formulations are introduced. Next, Section IV depicts the proposed solution methodology. Section V analyzes the experimental results, and conclusions are drawn in Section VI.

## II. MG AND EV MODELING

### A. MG MODEL

In microgrid scheduling, a probabilistic model should be given priority because of the small capacity and inherent renewable uncertainties in MGs. Among them, the PV output follows the Beta distribution, the WT output obeys the Weibull distribution, and the original load power obeys the normal distribution. Their respective probability models and probability density functions (PDFs) are detailedly given in the literature [2, 19].

### B. EV CHARGING MODEL

Electric vehicle charging model is the basis for investigating the orderly charging of EVs [27, 28]. In order to simplify the problem, this study only considers EV charging modes. And the main factors affecting the EV charging include the following two aspects:

#### 1) BATTERY SELECTION

There are many batteries can be used as candidates of EV batteries, such as lead-acid batteries, nickel-hydrogen batteries, lithium-ion batteries, and so on [1]. Each kind of battery has its technical and economic characteristics. Compared with other batteries, a lithium-ion battery has some obvious advantages, such as relatively longer service life, higher charging efficiency and greater depth of discharge [29]. Consequently, the lithium-ion battery is chosen as the EV battery in this work. In addition, the charging modes of lithium-ion batteries consist of slow charging and fast charging. Taking into account that fast charging modes are crucial for public acceptance of EVs, this study focuses on fast charging of lithium-ion batteries.

#### 2) TRAVEL NEEDS AND USAGE HABITS

It is assumed that there is no a parking space in charging stations in this study, and the default EV charging mode is that an EV leaves the charging station immediately once it is charged to its expected capacity. In this mode, the travel demand and usage habits mainly reflect in the EV arrival time, daily travel miles. These factors determine the total charging amount and the charging time for EV users. Based on the results of the 2009 National Traffic Survey of vehicles in the United States, the PDFs of the daily travel time and daily travel miles of EVs are obtained [30].

The EV arrival time obeys the normal distribution, and its PDF is

$$f_{IN}(t_{in}) = \begin{cases} \frac{1}{\sqrt{2\pi}\sigma_{in}} \exp[-\frac{(t_{in}+24-\mu_{in})^2}{2\sigma_{in}^2}], \\ \quad 0 < t_{in} \le \mu_{in} - 12 \\ \frac{1}{\sqrt{2\pi}\sigma_{in}} \exp[-\frac{(t_{in}-\mu_{in})^2}{2\sigma_{in}^2}], \\ \quad \mu_{in} - 12 < t_{in} \le 24 \end{cases} \quad (1)$$

where $t_{in}$ represents the EV arrival time, $\sigma_{in}$ and $\mu_{in}$ are the standard deviation and the mean value of $t_{in}$.

The daily travel mile of an electric vehicle is subjected to a normal distribution, and its PDF is

$$f_M(M_d) = \frac{1}{\sqrt{2\pi}\sigma_M M_d} \exp[-\frac{(\ln M_d - \mu_M)^2}{2\sigma_M^2}] \quad (2)$$

where $M_d$ represents the daily mileage of EVs, $\sigma_M$ and $\mu_M$ are the standard deviation and the mean value of $M_d$.

The model of SOC of EV batteries is given in this section. It can be seen from [31] that the initial SOC of an EV obeys a normal distribution when the EV arrives at the charging station. Based on the travel mileage of EV $i$ and its initial SOC, the actual SOC at the end of the charging can be calculated according to (3)-(5).

$$S_{i,real} = S_{i,s} + \frac{M_{i,d} E_{d100}}{100 B_{i,c}} \quad (3)$$

$$S_{i,e} \le S_{i,real} \le S_{i,max} \quad (4)$$

$$S_{i,min} \le S_{i,s} \le S_{i,e} \quad (5)$$

where $S_{i,e}$ and $S_{i,s}$ are respectively the expected SOC and the initial SOC of EV $i$; $S_{i,real}$ indicates the real state of charge; $S_{i,min}$ and $S_{i,max}$ are the minimum and maximum state of charge of EV $i$; $M_{i,d}$ denotes the travel miles of EV $i$; $E_{d100}$ is the power need when EV $i$ travels 100 kilometers; $B_{i,c}$ is the lithium battery capacity of EV $i$.

The charging time of EV $i$ can be calculated by

$$T_{i,CH} = \frac{(S_{i,real} - S_{i,s})B_{i,c}}{P_{i,rated}^{EV} \eta_{i,EV}^{CH}} \quad (6)$$

$$0 \le T_{i,CH} \le T_{i,max} \quad (7)$$

where $T_{i,CH}$ and $T_{i,max}$ denote the charging time and its maximum value of EV $i$; $P_{i,rated}^{EV}$ and $\eta_{i,EV}^{CH}$ are the rated electricity power and the charging efficiency of EV $i$.

## III. PROBLEM FORMULATION

In the proposed bi-level scheduling model, the upper level aims to minimize the net cost of the microgrid; the lower level seeks to the minimization of the EV charging cost. Focusing on coordinating renewable generation uncertainties and demand response and maintaining the dynamic supply-demand balance of power, a real-time pricing mechanism that acts as a bridge between the two levels is proposed in this work.

### A. THE UPPER LEVEL
1) OBJECTIVE FUNCTION

The upper level seeks to minimize the MG net operating cost, which is calculated by the difference between the operating cost and the revenue of the MG. Herein, the MG operating cost is the sum of both the fuel cost of microturbines (MTs) and the cost of spinning reserves provided by MTs and energy storage systems (ESSs) [2]. In this study, Zn-Br battery is chosen as the ESS since, compared with other batteries for grid-scale energy storage, it has many significant benefits, like lower costs, higher energy density, and longer service life.

The MG net operating cost can be expressed as follows:

$$\min F = -\sum_{t=1}^{T} P_t^{EV} \omega_{rt} + \left\{ \sum_{t=1}^{T} (g_1(P_t^{DC}) - g_2(P_t^{CH})) \right. \\ \left. + \sum_{t=1}^{T} \sum_{n=1}^{M_G} [(\varsigma_n R_{n,t}^{MT} + \kappa_n S_{n,t} + U_{n,t}(\zeta_n + \psi_n P_{n,t}^{MT}))] \right\} \quad (8)$$

where $t$ denotes a scheduling period (in hours) in an entire scheduling cycle $T$ ($T=24$ h in this study), $P_t^{EV}$ represents the power of the EV in period $t$, $P_t^{CH}$ and $P_t^{DC}$ respectively represent the charge-discharge power of the ESS in period $t$, $g_1(P_t^{CH})$ and $g_1(P_t^{DC})$ represent the charge and discharge costs of the ESS, respectively. $M_G$ is the total number of MT units, $\zeta_n$ and $\psi_n$ denote the consumption factors of the $n$th MT ($n \in M_G$), $\varsigma_n$ and $\kappa_n$ are the starting cost and the spinning reserve cost of MT $n$. $S_{n,t}$ and $U_{n,t}$ are the start-up variable and state variable of MT $n$. $R_{n,t}^{MT}$ and $P_{n,t}^{MT}$ are the spinning reserve and the output power provided by the MT.

2) CONSTRAINT CONDITIONS

In order to ensure the safe and stable operation of the system, the following constraints should be met:

$$U_{n,t} P_{n,min}^{MT} \le P_{n,t}^{MT} \le U_{n,t} P_{n,max}^{MT}, \forall t, n \in M_G \quad (9)$$

$$\sum_{n=1}^{M_G} P_{n,t}^{MT} + P_t^{DC} - P_t^{CH} + E_t = P_t^L + P_t^{EV} + P_t^{UN}, \forall t \quad (10)$$

$$SOC_{t+1} = \begin{cases} SOC_t + \eta^{CH} P_t^{CH} \Delta t \\ SOC_t - \Delta t P_t^{DC} / \eta^{DC} \end{cases}, \forall t \quad (11)$$

$$SOC_{min} \le SOC_t \le SOC_{max}, \forall t \quad (12)$$

$$\begin{cases} 0 \le P_t^{CH} \le P_{max}^{CH} \\ 0 \le P_t^{DC} \le P_{max}^{DC} \end{cases} \forall t \quad (13)$$

$$\begin{cases} 0 \le Q_t^{CH} \le Q_{max}^{CH} \\ 0 \le Q_t^{DC} \le Q_{max}^{DC} \end{cases} \forall t \quad (14)$$

$$V_{min} \le V_t \le V_{max}, \forall t \quad (15)$$

$$SOC_0 = SOC_{Tend} = SOC_* \tag{16}$$

$$P_{n,t}^{MT} + R_{n,t}^{MT} \leq U_{n,t} P_{n,\max}^{MT}, \forall t, n \in M_G \tag{17}$$

$$P_{Ress,t} \leq \min\{\eta^{DC}(SOC_t - SOC_{\min})/\Delta t, P_{\max}^{DC} - P_t^{DC}\}, \forall t \tag{18}$$

$$P_{rob}\left\{\sum_{n=1}^{M_G} R_{n,t}^{MT} + P_{Ress,t} \geq E_t - (P_t^{WT} + P_t^{PV})\right\} \geq \gamma, \forall t \tag{19}$$

Eqs. (9) and (10) respectively represent the MT output constraint and the system power balance constraint. Here, $P_{n,\max}^{MT}$ and $P_{n,\min}^{MT}$ are the maximum and minimum power outputs of the $n$th MT unit; $P_t^L$ denotes the predicted active power of the original load; $P_t^{UN}$ is the power of the controlled load in period $t$; $E_t$ represents the expected value of renewable generations in period $t$.

Eqs. (11) - (16) represents the constraints of the ESS. Eq. (11) is the charge and discharge equation [2, 19, 32], where $\eta^{CH}$ and $\eta^{DC}$ are respectively the charge and discharge efficiencies. $\Delta t$ represents the duration of a time period (it is taken as 1 hour here). $SOC_t$ and $SOC_{t+1}$ are the energy stored in the ESS in period $t$ and $t+1$, respectively. Eq. (12) describes the capacity constraint of Zn-Br batteries in ESS, where $SOC_{\min}$ and $SOC_{\max}$ are the minimum and the maximum energy stored in the ESS. Eqs. (13) - (14) represents the charge and discharge rate constraints of ESS, where $P_{\max}^{CH}$ and $P_{\max}^{DC}$ are the maximum charge and discharge active powers of the ESS. $Q_{\max}^{CH}$ and $Q_{\max}^{DC}$ are the maximum charge and discharge reactive powers of the ESS. Eq. (15) is the voltage constraint, where $V_{\min}$ and $V_{\max}$ are the minimum and the maximum voltage of the battery. $V_t$ is the battery voltage in period $t$. Eq. (16) denotes the starting and ending constraint. In order to balance the energy stored in the ESS and prolong the battery life, the initial energy stored and the remaining energy at the end of a scheduling cycle should be equal [2, 33]. $SOC_*$ represents the initially stored energy limit of the ESS, $SOC_0$ represents the initial energy in the ESS, $Tend$ denotes the end of the total scheduling cycle (it is set to 24h here).

Eq. (17) - (19) represents the spinning reserve constraint of the microgrid. Since the main power grid does not supply power to the IMG, the spinning reserve is a significant resource for balancing both supply and demand sides [3, 32]. Eqs. (17) and (18) are the spinning reserve constraints for MT and ESS, respectively, where $P_{Ress,t}$ represents the reserve capacities of the ESS in period $t$. Considering that the joint output of renewable generations may be zero, in this small probability situation, the adequate spinning reserve must be provided to maintain the reliability of the system, but it will incur additional costs. Eq. (19) illustrates the probabilistic spinning reserve requirement, in which $\gamma$ denotes the confidence level.

### B. THE LOWER LEVEL
Electric vehicles have been recently receiving increasing attentions since they play a critical role in energy conservation and emission reduction. In this work, the lower level seeks to minimize the EV charging cost.

#### 1) OBJECTIVE FUNCTION
The EV charging cost is calculated by the following formula:

$$\min F_2 = \omega_{rt,t} \sum_{t=1}^{T} P_t^{EV} + \frac{W_c}{365 \cdot m} \tag{20}$$

where $\omega_{rt,t}$ is the real-time electricity price in period $t$, $P_t^{EV}$ represents the power of the EV in period $t$, $W_c$ denotes the investment cost of the charging station, and $m$ denotes the service life of the charging station.

#### 2) CONSTRAINT CONDITIONS
At any period, the EV charging power should not be greater than the maximum allowable power of the microgrid.

$$P_t^{EV} \leq \alpha \times (\sum_{n=1}^{M_G} P_{n,t}^{MT} + P_t^{DC} - P_t^{CH} - P_t^L) \tag{21}$$

where $P_t^L$ is the active power of the load in period $t$, $P_t^{EV}$ represents the EV charging power in period $t$, $\alpha$ is a regulatory factor that controls the upper limit of $P_t^{EV}$.

The charging power of EV $i$ should not exceed its upper and lower limits, which is given by

$$P_{i,t,\min}^{EV} \leq P_{i,t}^{EV} \leq P_{i,t,\max}^{EV}, \forall t \tag{22}$$

where $P_{i,t,\max}^{EV}$ and $P_{i,t,\min}^{EV}$ are respectively the maximum and minimum charging power of EV $i$ in period $t$.

The capacity of EV $i$ should be within a proper range, i.e. it should not be less than the EV users' expected capacity but not greater than the battery rated capacity, which is formulated as

$$S_{i,e} B_{i,c} \leq S_{i,s} B_{i,c} + P_{i,rated}^{EV} \eta_{i,EV}^{CH} T_{i,CH} \leq B_{i,c} \tag{23}$$

### C. REAL-TIME PRICING MECHANISM
In order to reflect the dynamic relationship between supply and demand, a real-time pricing mechanism is put forward [19] in this work. The main steps of this mechanism are shown as follows:

(a) The EV charging plan $P_t^{EV}$ is firstly obtained by solving the lower-level model, and then, the sum of $P_t^{EV}$ and the original load power $P_t^L$ are fed back to the upper-level model.

(b) The real-time electricity price is calculated according to the following formula in the upper level:

$$\omega_{rt,t} = \frac{\sum_{i=1}^{I} P_{i,t}^{EV} + P_t^L}{P_{REF}^L} \times \omega_{REF} \tag{24}$$

where $P_{REF}^L$ is the reference power of the original load, and $I$ is the total number of EVs; $\omega_{REF}$ represents the reference electricity price, and $\omega_{rt,t}$ denotes the real-time electricity price.

## IV. PROPOSED SOLUTION METHODOLOGY

Considering that a bi-level programming model is non-deterministic polynomial-time hard (NP-hard), a hybrid solution algorithm called JAYA-IPM is developed to ensure sufficient optimality and high computation efficiency. The algorithm solves the model through an iterative process between levels, and the optimal scheduling strategy is finally determined.

### A. SEQUENCE DESCRIPTION OF RENEWABLE GENERATIONS

The sequence operation theory (SOT) is here utilized to obtain probabilistic sequences of renewable generations, and then transform a chance constraint into its deterministic equivalent class, which avoids tedious and time-consuming Monte Carlo simulations in the solution process.

In this study, all renewable outputs are modelled by probabilistic sequences obtained through discretizing continuous probability distributions. Concretely speaking, PV and WT outputs are depicted via probabilistic sequences $a(i_{a,t})$ with length $N_{a,t}$ and $b(i_{b,t})$ with length $N_{b,t}$, which are defined as

$$a(i_{a,t}) = \begin{cases} \int_0^{q/2} f_P(P^{PV}) dP^{PV}, & i_{a,t} = 0 \\ \int_{i_{a,t}q-q/2}^{i_{a,t}q+q/2} f_P(P^{PV}) dP^{PV}, & i_{a,t} > 0, i_{a,t} \neq N_{a,t} \\ \int_{i_{a,t}q-q/2}^{i_{a,t}q} f_P(P^{PV}) dP^{PV}, & i_{a,t} = N_{a,t} \end{cases} \quad (25)$$

$$b(i_{b,t}) = \begin{cases} \int_0^{q/2} f_0(P^{WT}) dP^{WT}, & i_{b,t} = 0 \\ \int_{i_{b,t}q-q/2}^{i_{b,t}q+q/2} f_0(P^{WT}) dP^{WT}, & i_{b,t} > 0, i_{b,t} \neq N_{b,t} \\ \int_{i_{b,t}q-q/2}^{i_{b,t}q} f_P(P^{WT}) dP^{WT}, & i_{b,t} = N_{b,t} \end{cases} \quad (26)$$

### B. HANDLING OF CHANCE CONSTRAINTS

1) PROBABILISTIC SEQUENCES OF RENEWABLE GENERATIONS

The probability sequence $c(i_{c,t})$ of the joint power outputs is obtained according to the addition-type-convolution (ATC) operation in the SOT [34]:

$$c(i_{c,t}) = a(i_{a,t}) \oplus b(i_{b,t}) = \sum_{i_{a,t}+i_{b,t}=i_{c,t}} a(i_{a,t}) b(i_{b,t}), \quad i_{c,t} = 0,1,...,N_{a,t}+N_{b,t} \quad (27)$$

The joint power output and its probabilistic sequence are illustrated in Table 1.

**Table 1** Joint power output and its probabilistic sequence

| Power (kW) | 0 | $q$ | … | $u_c q$ | … | $N_{c,t} q$ |
|---|---|---|---|---|---|---|
| Probability | $c(0)$ | $c(1)$ | … | $c(u_c)$ | … | $c(N_{c,t})$ |

2) DETERMINISTIC TRANSFORMATION OF CHANCE CONSTRAINTS

To handle the chance constraint in (19), a new 0-1 variable $W_{u_{c,t}}$ is defined as [2, 19]:

$$W_{u_{c,t}} = \begin{cases} 1, & \sum_{n=1}^{M_G} R_{n,t}^{MT} + P_{Ress,t} \geq E_t - u_{c,t} q \\ 0, & \text{otherwise} \end{cases} \quad (28)$$

$$\forall t, u_{c,t} = 0,1,...,N_{c,t}$$

From (28), it can be seen that if the total spinning reserve $\sum_{n=1}^{M_G} R_{n,t}^{MT} + P_{Ress,t}$ is not less than the difference between $E_t$ and $u_{c,t} q$, $W_{u_{c,t}}$ is set to 1; otherwise, it is 0.

According to Table 1, $c(u_c)$ is the probability that corresponds to the joint power output $u_c q$. Thereby, by submitting (28) into (19), Eq. (19) can be rewritten as follows:

$$\sum_{u_{c,t}=0}^{N_{c,t}} W_{u_{c,t}} c(u_{c,t}) \geq \gamma, \quad \forall t \quad (29)$$

### C. JAYA ALGORITHM

JAYA is a powerful algorithm to handle complex optimization issues, which is proposed by R. Venkata Rao. Since it requires no algorithm-specific parameter expect two common control parameters namely the population size and maximum number of iterations, its results are more stable than other intelligent optimization algorithms [19].

1) BASIC PRINCIPLES

The key idea of the Jaya algorithm is that a solution must move away from the worst solution and move to the best solution [35]. If $X_{j,k,i}$ is the value of variable $j$ for candidate $k$ at iteration $i$, then $X_{j,k,i}$ is calculated as

$$X'_{j,k,i} = X_{j,k,i} + r_{1,j,i}(X_{j,best,i} - |X_{j,k,i}|) \\ \times r_{2,j,i}(X_{j,worst,i} - |X_{j,k,i}|) \quad (30)$$

where $X_{j,best,i}$ (/ $X_{j,worst,i}$) is the value of the variable $j$ for the best (/worst) candidate; $X'_{j,k,i}$ is the updated value of $X_{j,k,i}$; $r_{1,j,i}$ and $r_{2,j,i}$ are two random numbers for variable $j$ at iteration $i$; $X'_{j,k,i}$ is accepted if it generates a better objective value.

2) HYBRID CODING

To speed up the optimization process, this works utilizes a hybrid real/integer-coded scheme [36]. The used variables are classified into two classes: the continuous variables $P_n^{MT}$, $R_n^{MT}$, $P_{Ress}$, $P^{EL}$, $\omega_{rt}$, $P^{CH}$, $P^{DC}$ and the discrete variables $U_n$ and $S_n$.

3) PREVENTION OF PREMATURE CONVERGENCE

In order to avoid local minima and premature convergence, this paper adopts a disturbance optimization strategy. First, a dynamic evolutionary monitoring mechanism is used through analyzing the fitness variance of the population during the optimization process. Once a convergence criterion is satisfied, the disturbance optimization strategy is implemented. Concretely speaking, a disturbance is added by re-initializing a certain percentage of individuals that are randomly selected. The used convergence criterion is

$$thr1 < \sigma_{i+1}^2 / \sigma_i^2 < thr2 \qquad (31)$$

where, $\sigma_{i+1}^2$ and $\sigma_i^2$ are respectively the fitness variance at the ($i$+1)th and $i$th iterations. Here, the monitoring thresholds thr1 and thr2 are set to 0.99 and 1.01.

### D. INTERIOR POINT METHOD
The interior point method originally proposed by John von Neumann is a classical optimization method for addressing linear programming [37], and its key principle is to gradually approximate the optimal solution of the original problem in the feasible domain. In view of the IPM's advantages of high efficiency and excellent accuracies, it is used to solve the lower-level model.

### E. DETERMINATION OF THE OPTIMAL SCHEME
In this paper, it is necessary to introduce a joint optimization function $F^{JO}$ to screen out the optimal joint operation schemes from the yielded candidate ones during iterations [19].

$$F^{JO} = \min \sqrt{\left(F_1^{JO} - F_1^{IO}\right)^2 + \left(F_2^{JO} - F_2^{IO}\right)^2} \qquad (32)$$

where $F_1^{IO}$ and $F_2^{IO}$ are respectively the IMG operating cost without considering the EV interests and the EV charging cost without considering the MG revenues; while $F_1^{JO}$ and $F_2^{JO}$ are respectively the operation cost of the IMG and the EV charging cost under consideration of demand response. When this objective function takes the minimum value during iterations, the scheduling scheme corresponding to this iteration is chosen as the optimal scheme [19]. In this case, the resulting optimal scheduling scheme is capable of balancing the interests of the microgrid and electric vehicle users, achieving a win-win situation for both.

### F. SOLVING PROCESS
Fig. 1 shows the procedure of the proposed solution method, and the specific steps are as follows:

Step 1: Build the IMG model according to (8) ~ (19).
Step 2: Convert the chance constraint into its deterministic equivalence class.
Step 3: Set the MG parameters.
Step 4: Set the electricity price of the main grid as the basic price of the MG before optimization. The price is known in advance on the basis of historical data, the load demand of previous day/hour and the expected load demand of next hour or day.
Step 5: Calculate the real-time electricity price according to formula (24).
Step 6: Optimize the upper-level model by the JAYA algorithm.
Step 7: Obtain the MG optimal scheduling scheme.
Step 8: Build the lower-level model according to (20) ~ (23).
Step 9: Solve the lower-level model by using the IPM.

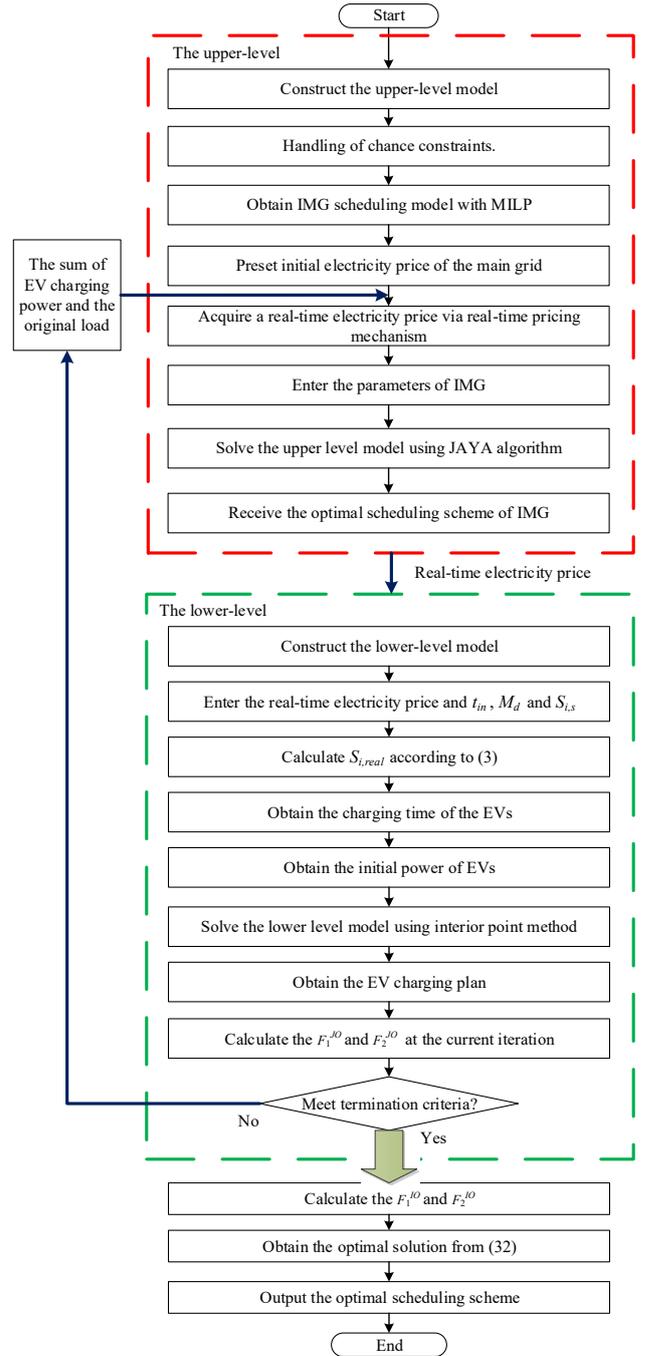

FIGURE 1. Flowchart of the proposed hybrid solution approach

Step 10: Obtain the EV optimal charging plan.
Step 11: Calculate the MG net cost $F_1^{JO}$ and the user cost $F_2^{JO}$ at the current iteration.
Step 12: Judge whether the termination criteria is met. If met, end the optimization process and proceed to the next step, otherwise, return to step 5. Here, the used termination criterion is whether the current iteration exceeds the pre-defined maximum number of iterations.
Step13: Calculate the costs $F_1^{IO}$ and $F_2^{IO}$.
Step14: Identity the optimal solution with the minimum value of $F^{JO}$ according to (32).
Step15: Obtain the optimal scheduling scheme.

## V. CASE STUDY

The proposed approach has been examined on an improved MG testing system, which is illustrated in Fig. 2. This system consists of one PV control board, three MT units, one WT unit, an EV charging station, and the original load. Among them, PCC denotes a common coupling point.

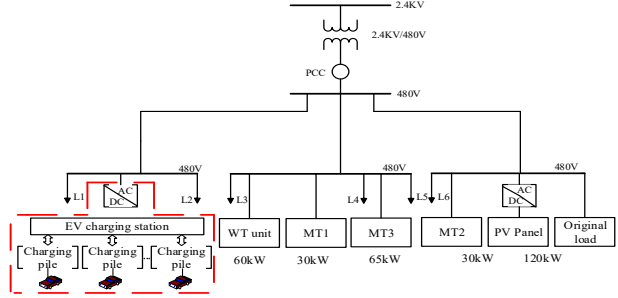

FIGURE 2. Microgrid test system

### A. PARAMETER SETTINGS
1) MG PARAMETERS
Table 2 gives the used parameters of MT units in this work.

Table 2 Parameters of MT units

| MT number | $\zeta$($) | $\kappa_n$ ($) | $\Psi$($/kW) | $\varsigma$($/kW) | $P_{min}^{MT}$ (kW) | $P_{max}^{MT}$ (kW) |
|---|---|---|---|---|---|---|
| MT1 | 1.2 | 1.6 | 0.35 | 0.04 | 5 | 35 |
| MT2 | 1.2 | 1.6 | 0.35 | 0.04 | 5 | 30 |
| MT3 | 1.0 | 3.5 | 0.26 | 0.04 | 10 | 65 |

In the above table, $P_{max}^{MT}$ and $P_{min}^{MT}$ are the maximum and minimum values of the MT outputs; the parameters of the Zn-Br battery are as follows: $P_{max}^{DC} = P_{max}^{CH}$ =40 kW, $\eta^{DC} = \eta^{CH}$ =0.95, $SOC_{max}$ =160 kWh, $SOC_{min}$ =32 kWh; the ESS reserve cost is $\omega_{rc}$ =0.02 $/kW and 0.3 $/kWh and 0.5 $/kWh are respectively the charge/discharge prices of ESS [2]; the reference power $P_{REF}^L$ and the maximum power of the original load $P_{max}^L$ are respectively 80.00 kW and 57.26 kW [19], and the used reference price $\omega_{REF}$ is 0.6 $/kWh, $N_{iter,max}$ is set to 20.

2) EV PARAMETERS
In this study, the used EV parameters are as follows: the rated capacity $B_{i,c}$, the rated charging power $P_{i,rated}^{EV}$ and the charging efficiency $\eta_{i,EV}^{CH}$ of EV $i$ are respectively 19 kWh, 7.5 kW and 0.95; the power consumption per 100 kilometers of an EV is set to $E_{d100}$ =15kWh, the regulatory factor $\alpha$ is 0.4, and the total number of EVs $I$ is set to 20; $S_{i,min}$ and $S_{i,max}$ are respectively 0.2 and 1; $\mu_{in}$ =17.47; $\sigma_{in}$ =3.41; $\mu_M$ =40; $\sigma_M$ =15. Note that, it is assumed that the EV expected capacity is 90%; the investment cost and the service life of a charging pile are respectively $ 3000 USD and 10 years.

3) ALGORITHM PARAMETERS
The parameters of the Jaya are set as follows: the population size and the maximum number of iterations are respectively 100 and 1500. Other algorithm parameters are assigned as follows: the load fluctuation $\partial$ =10%, the confidence level $\gamma$ =95% and the step size $q$=2.5 kW.

### B. BASIC DATA
This section give the outputs of renewable generations, and the EV arrival and departure times. All of them are used as the basic data for the subsequent analysis. The original load power and renewable outputs are illustrated in Fig. 3.

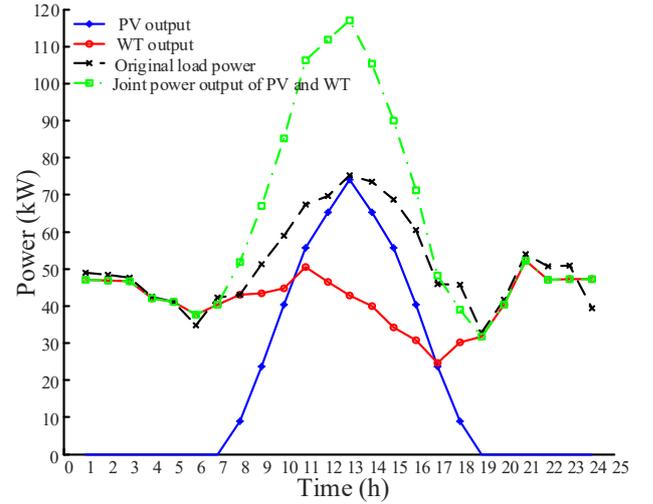

FIGURE 3. Load and the renewable outputs

Fig. 3 shows that the original load and the joint output of renewable generations obtained via the ATC operation between WT and PV power outputs. By doing so, the uncertainties of multiple renewable generations are effectively tackled via the SOT [34].

Based on the results of the national household travel survey in 2009 made by the Federal Highway Administration of the Department of Transportation of the United States, an EV arrival case following the normal distribution in an entire scheduling period is randomly taken as an example in this study. The EV arrival and departure times at the EV charging station are illustrated in Fig. 4.

### C. ECONOMIC COST ANALYSIS
For purpose of examining the economy of the joint optimization between IMG and EV, three different strategies are designed in this paper.
  **Strategy 1:** IMG scheduling without consideration of the EV charging costs;
  **Strategy 2:** Joint optimization of IMG and EV with consideration of demand response.
  **Strategy 3:** EV scheduling without consideration of the MG revenues.
  The MG net operating costs and the EV charging costs in the above strategies are shown in Fig. 5.

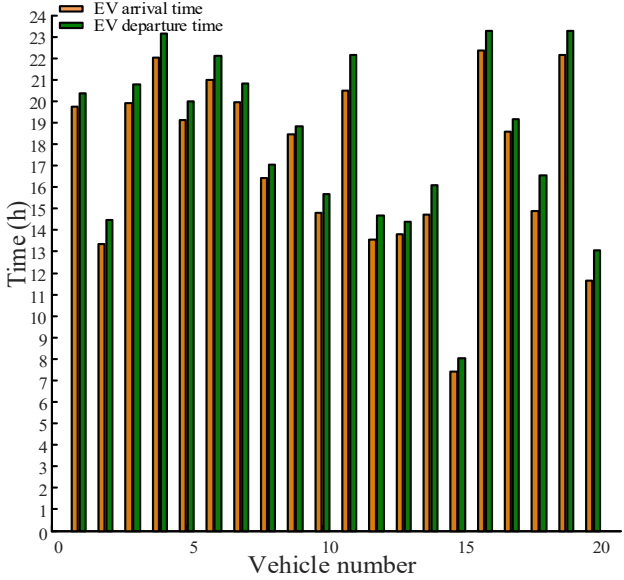

**FIGURE 4.** EV arrival and departure times

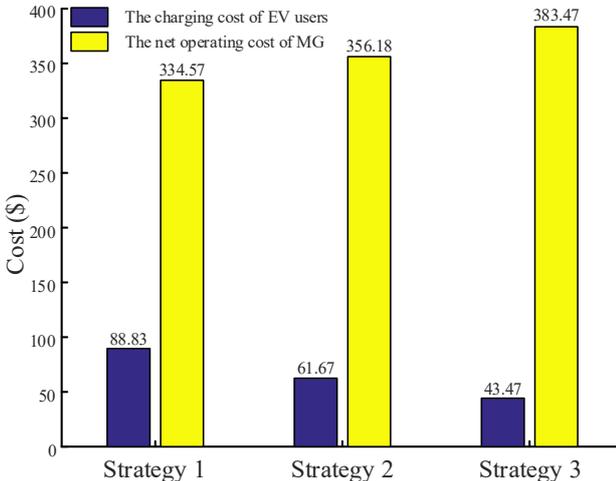

**FIGURE 5.** IMG net operating cost and the charging cost of EVs in different strategies

Fig. 5 illustrates that the strategy 2 is the best one among the three strategies since it can balance the interests of MG and EV users; while the other strategies only consider the interests of MG or EV users, separately. In particular, regarding the EV charging cost of the IMG, the result in strategy 2 is superior to that in strategy 1; while as far as the net operating cost of the IMG is concerned, strategy 2 outperforms strategy 3. On the basis of this phenomenon, it can be concluded that demand response of electric vehicles is able to improve the economy of IMG and EV users, which plays an important role in the scheduling of isolated microgrids with renewables.

The electricity prices of the main grid in this work are listed in Table 3.

**Table 3** The electricity prices of the power grid

| Periods | Specific time periods | Price ($/kWh) |
|---|---|---|
| Peak period | 11:00-15:00 | 0.83 |
| Flat period | 00:00-06:00, 07:00-11:00,15:00-18:00,19:00-24:00 | 0.62 |
| Off-peak period | 00:60-07:00,18:00-19:00 | 0.17 |

The real-time electricity prices and the sum of both the initial powers of EVs and the original load are demonstrated in Fig. 6.

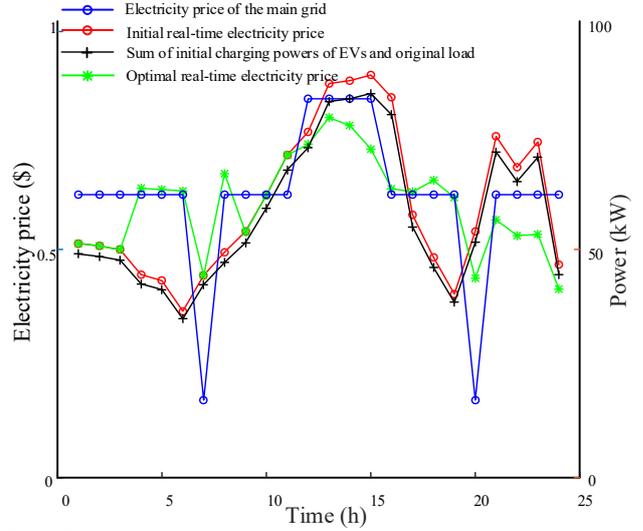

**FIGURE 6.** Electricity price curves

Fig. 6 shows that the proposed real-time electricity prices (including the initial and optimal real-time prices) significantly outperform the electricity price of the main grid, since the latter is unavailable to adjust the electricity prices according to the dynamic supply-demand relationships. Furthermore, the optimal real-time prices are superior to the initial real-time prices, since the former is able to reduce the peak-to-valley difference of load powers.

### D. RESERVE CAPACITIES UNDER DIFFERENT CONFIDENCE LEVELS

This section discusses the spinning reserve capacities under different confidence levels. Spinning reserve is an important auxiliary service for balancing source-load difference [2], and confidence levels $\gamma$ determine the reserve capacities of MGs. The relationship between reserve capacity and confidence level is shown in Fig. 7.

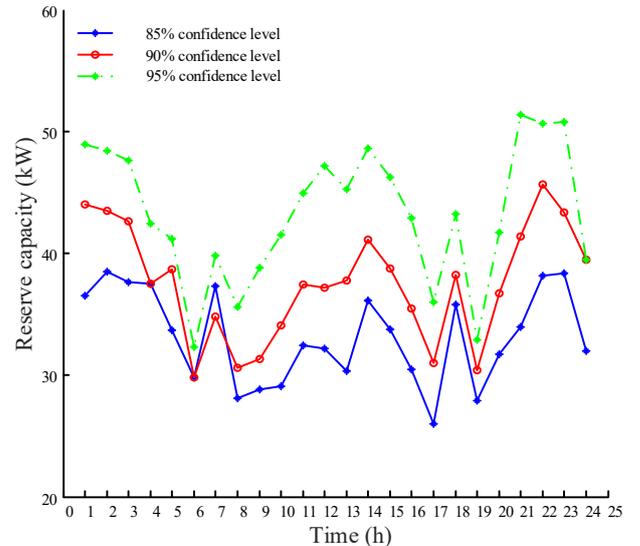

**FIGURE 7.** Reserve capacities under different confidence levels

Fig. 7 indicates that the MG reserve capacities are dependent on confidence levels. It's known that reserve capacities are closely related to the economy and reliability of the MG operation. For one thing, a higher confidence level can increase the system reliability at the cost of the economy; for another, a less confidence level will bring out a decreased reliability and a better economy. As a result, a suitable confidence level is crucial to trade off the reliability and economy of the MG operation.

### E. IMPACTS OF DEMAND RESPONSE
For purpose of properly evaluating the performances of the PBDR strategy, the following two cases are designed:

**Case 1**——without considering demand response: the charging price of EVs is set to the price of the main grid.

**Case 2**——considering demand response: the charging price of EV adopts real-time electricity prices.

#### 1) IMPACT ON MG OPERATION
Comparative tests without and with consideration of demand response have been performed, and the test results are respectively demonstrated in Figs. 8 and 9.

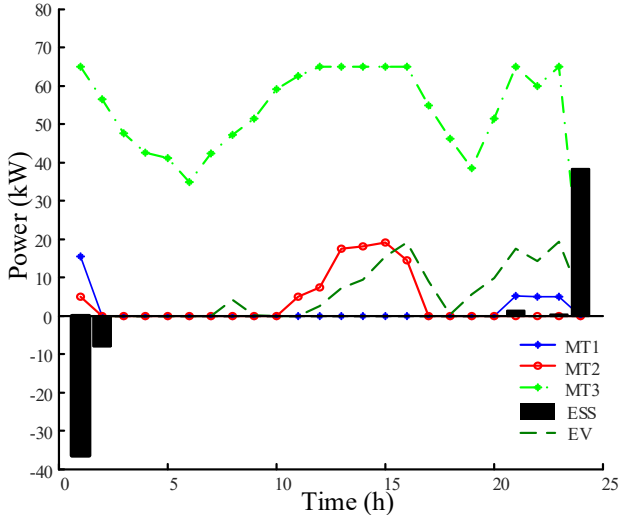

**FIGURE 8.** Scheduling strategy in case 1

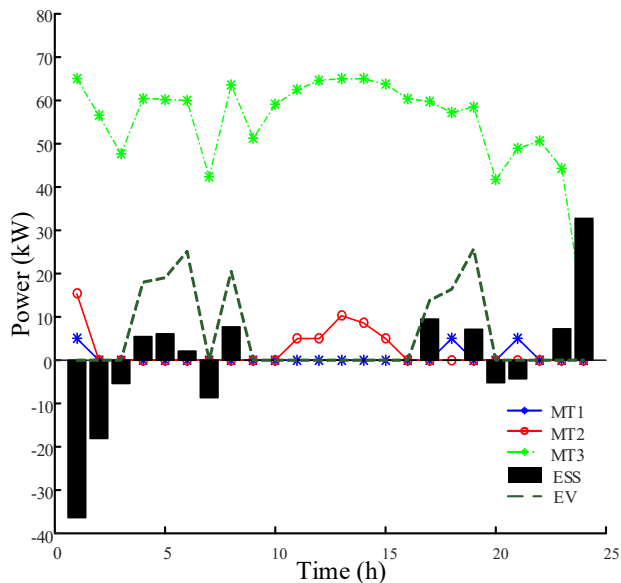

**FIGURE 9.** Scheduling strategy in case 2

It can be seen from Figs. 8 and 9 that demand response manages to promote the ESS and EV users actively participating in economic operations of the microgrid. (1) Regarding the ESS, the ESS charging-discharging frequencies in case 2 are obviously greater than those in case 1. This is because the ESS will absorb more electrical energy to suppress load fluctuations when its discharging price is higher than the charging price. (2) In terms of EVs, due to the introduction of PBDR, the charging behaviors of EVs can flexibly response to the changes of electricity prices. Therefore, ones can conclude that PBDR is an effective way for guiding the ESS and EV users actively participate in the microgrid optimal scheduling.

#### 2) IMPACT ON EV USERS
The EV charging powers without and with consideration of demand response are illustrated in Fig. 10.

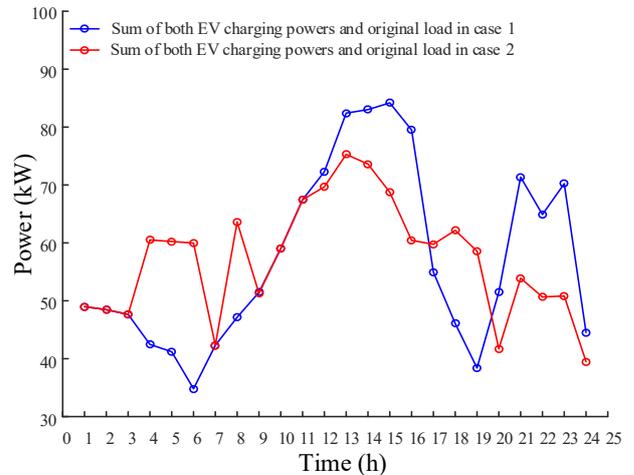

**FIGURE 10.** EV Load powers without and with considering demand response

From Fig. 10, it can be observed that demand response plays a key role in guiding the charging behaviors of EV users. Concretely speaking, EVs users decrease the charging powers in the peak-load periods, while they increase the charging powers in off-peak periods. By this means, the peak load shaving is achieved by leveraging EV flexibility on the load side while maintaining the source-load balance, thereby promoting EV users to actively participate in the MG scheduling. By this means, it provides a fundamental way to balance the interests of both MG and users.

## VI. CONCLUSION
This paper proposes a bi-level programming model for scheduling of isolated microgrids with renewable generations by incorporating demand response of electric vehicles. And thereby, a hybrid solution algorithm JAYA-IPM is developed to solve the model. The simulation results demonstrate that demand response of electric vehicles is able to guide EV users to actively participate in MG scheduling and achieve the peak load shaving, which offers a "win-win" solution to balance the interests between microgrid and EV users.

Future work will focus on extending the proposed approach to scheduling of heat and electricity integrated energy system. Note that it is assumed in this paper that only charging modes are available for EVs, while a more realistic scenario shall consider EV discharging modes. Besides, the charging waiting time of EV users in this paper is ignored, while it is necessary for real-world applications due to limited charging piles.